\documentstyle[prd,aps,floats,tabularx,epsfig]{revtex}
\tighten

\newcommand{\al}{\alpha}

\newcommand{\de}{\delta}

\newcommand{\La}{\Lambda}
\newcommand{\la}{\lambda}
\newcommand{\Om}{\Omega}

\newcommand{\si}{\sigma}

\newcommand{\th}{\theta}

\newcommand{\DD}{\mbox{$\cal D$}}
\newcommand{\GG}{\mbox{$\cal G$}}
\newcommand{\SS}{\mbox{$\cal S$}}

\newcommand{\Eq}[1]{ Eq.~(\ref{#1})}
\newcommand{\be}{\begin{equation}}
\newcommand{\ee}{\end{equation}}

\newcommand{\bea}{\begin{eqnarray}}
\newcommand{\eea}{\end{eqnarray}}
\newcommand{\bean}{\begin{eqnarray*}}
\newcommand{\eean}{\end{eqnarray*}}
\newcommand{\dd}{\partial}

\newcommand{\bk}{{\bf k}}

\def\spose#1{\hbox to 0pt{#1\hss}} 
\def\ltapprox{\mathrel{\spose{\lower 3pt\hbox{$\mathchar"218$}} 
 \raise 2.0pt\hbox{$\mathchar"13C$}}} 
\def\gtapprox{\mathrel{\spose{\lower 3pt\hbox{$\mathchar"218$}} 
 \raise 2.0pt\hbox{$\mathchar"13E$}}} 
\def\inapprox{\mathrel{\spose{\lower 3pt\hbox{$\mathchar"218$}} 
 \raise 2.0pt\hbox{$\mathchar"232$}}} 
 
\begin{document}
\draft
\preprint{\ 
}

\twocolumn[\hsize\textwidth\columnwidth\hsize\csname@twocolumnfalse\endcsname 
\title{Cosmic microwave background anisotropies and extra dimensions 
	in string cosmology}
\author{ A. Melchiorri$^{1,3}$, F. Vernizzi$^1$, R. Durrer$^1$ and 
	G. Veneziano$^{2}$ }
\address{ $^1$D\'epartement de Physique Th\'eorique, Universit\'e de Gen\`eve,
24 quai Ernest Ansermet, CH-1211 Gen\`eve 4, Switzerland\\
$^2$CERN Theory Division, CH-1211, Gen\`eve 23, Switzerland\\
$^3$Dipartimento di Fisica, Universit\'a di Roma Tor Vergata, Via della Ricerca Scientifica, Roma, I-00133, Italy}
\maketitle

\begin{abstract}

A recently proposed mechanism for large-scale structure in string 
cosmology --based on massless axionic seeds-- is further
analyzed and extended to the acoustic-peak region.
 Existence, structure, and height of the peaks turn out to 
depend crucially on the overall evolution of extra dimensions
during the pre-big bang phase: conversely,  precise cosmic microwave
background  anisotropy data  in the acoustic-peak region will provide
a window on string-theory's extra dimensions 
 before their eventual compactification.

\end{abstract}

\date{\today}

] 
{

 One of the most stringent tests of
inflationary cosmology will come when new precise 
satellite data on cosmic microwave background (CMB) anisotropies
down to small angular scales will become available
 during the next years \cite{MAP}.
 Hopefully, these data will allow not only
to check whether the generic paradigm of inflation is valid, 
but also to make a strong selection among the multitude of models
of inflation which are presently on the market.  Models
differ, in particular, on the presence or absence of a sizeable
tensor component (to be detected by polarization experiments),
on the possible non-Gaussianity of the fluctuations (to be tested
through higher-order correlations) and, finally, on the height,
and position of the so-called acoustic peaks in the  multipole
coefficients $C_{\ell}$ in the region $\ell > 100$.

The pre-big bang (PBB) scenario \cite{PBB}, a particular model of inflation
 inspired by the duality properties
of string theory, was thought for sometime to be unable to 
provide a quasi-scale-invariant (Harrison-Zeldovich, HZ) spectrum of 
perturbations. Indeed,  first-order tensor and scalar perturbations
 were found  to be characterized by
 extremely `blue' spectra~\cite{PBB}. The large tilt, together with 
 a natural normalization imposed by the string cut-off 
 at the shortest amplified scales ($\sim 1$mm), makes their contribution
to large-scale structure completely negligible. 

It was later realized \cite{Copeland}, however, that the spectral tilt of
the supersymmetric partner of the dilaton, the  universal
axion of string theory (not to be confused with the Peccei-Quinn
axion!), $\sigma$, can have a whole range
 of values, depending on the 
overall behaviour of the six compactified internal dimensions.
It is most useful to express the result
in terms of the axion energy spectrum
during the radiation era \cite{Buon,Hadad}.
Let us define the tilt  $\alpha$ by:
\be
\Omega_{\si}(k,\eta) \equiv \rho_c^{-1}~ d \rho_{\si}(k, \eta) / d \log k 
 \propto (k/k_1)^{\alpha} ~,
\ee
where, as usual, $\rho_c$ is the critical energy density, and $k_1$, related to
the string scale, represents the end-point of the spectrum.
Assuming, as an example, separate isotropic behaviour for the three
external and the six internal dimensions, one finds:
\be
 \alpha =  {3 + 3 r^2 - 2 \sqrt{3 + 6 r^2} \over 1 + 3 r^2} ~,
\label{spectralindex}  
\ee
where $r \equiv {1\over 2} (\dot{V_6}~V_3)/(V_6~\dot{V_3})$
 is a measure
of the relative evolution of the internal and external volumes.

Eq. (\ref{spectralindex}) allows for  a range of values for 
the tilt $\alpha$. For static internal dimensions ($r=0$) one finds
a negative tilt, a `red' spectrum with 
$\alpha = 3 - 2 \sqrt3 \sim -.46$; for static
external dimensions ($r = \infty$) one finds a `blue' spectrum 
with $\alpha = 1$ while, finally, for a
globally isotropic evolution (modulo T-duality), i.e. for $r= \pm 1$,
one obtains a flat HZ spectrum, $\alpha = 0$ \cite{Buon}.
As we shall show in this paper, CMB anisotropy data prefer a slightly
blue spectrum with $\al\sim 0.4$ leading to $r\sim 2.2$ so that the
internal dimensions contract somewhat faster than the external
dimensions expand.
 We note also that the pure
power-law behaviour in (\ref{spectralindex}) is only valid if PBB evolution
is not itself composed of various phases: it is conceivable, {\em e.g.},
that some of the internal dimensions  may `freeze'
sometime during the PBB phase, in which case  $\alpha$ will undergo a 
(negative) jump at some characteristic scale $k^*$ related to the
freeze-out time. We will come to this possibility below.

The results of \cite{Copeland,Buon,Hadad}
 reopened the possibility that PBB cosmology may 
contain a natural mechanism for generating
large-scale anisotropy via the 'seed' mechanism \cite{d90}. 
This possibility, which
belongs to the generic class of isocurvature perturbations, 
is analyzed in \cite{1} for massless axions, to which
we shall limit our attention in this letter, and in \cite{3}
for very light axions. Isocurvature perturbations from scalar fields
have also been discussed in Ref.\cite{isocurvature}, but there the
scalar field perturbations just determine the initial conditions.
In our model the axion pays the role of a 'seed' like in scenarios
 with topological defects. The power spectrum of the seed is however
 not determined by causality, but the spectral index can vary (within
the above limits). This reflects the fact that the axion field is
 generated during an inflationary phase.

In the above papers a strong correlation between the
tilt (the value of $n_s - 1$ in standard notations)
 and normalization of the $C_{\ell}$'s was noticed.
 A range of values  around
$n_s = 1.2$ (slightly blue spectra) 
appeared to be favored by a simultaneous fit to
the tilt and normalization on the large angular scales
 observed by
 COBE \cite{COBE} to which  
the analysis in \cite{1} was actually confined. 
In this paper we extent this study down to the small angular scales
which have been explored observationally with limited precision so 
far \cite{Teg} but which will become
precisely determined during the next decade. We also supplement the analytic
study of \cite{1} with numerical calculations.

As in previous work \cite{1,3} we suppose that 
the contribution of the axions to the cosmic fluid can
be neglected and that they interact with it only
gravitationally. They then play the role of 'seeds' which generate 
fluctuations in the cosmic fluid\cite{d90}.

The  evolution of axion perturbations is determined by the
well-known axion-free background of string cosmology. One finds~\cite{1}
\be
 \ddot{\psi} +\left(k^2-{\ddot{a}_A\over a_{A}}\right)\psi =0 ~, \label{evol}
\ee
where we have introduced the `canonical' axion field 
$\psi=a_{A}\si$. The function $a_{A}=ae^{\phi/2}$ is the axion pump field, 
$a$ denotes
the scale factor in the string frame, and $\phi$  is the dilaton,  which is
supposed to be frozen after the pre-big bang/post-big bang
transition. Dots denote derivation w.r.t. conformal time $\eta$.
The initial condition for  \Eq{evol} is obtained from the  pre-big bang
solution and is then evolved numerically with  $a_{A}=a$ during the
post big bang. The pre-big bang initial conditions require~\cite{1}
\be
\si (\bk, \eta) = {c (\bk)\over a\sqrt k} \varphi( k,\eta) , ~
\varphi(k,\eta) = \sin k\eta , ~~\eta\ll \eta_{eq}.
\ee
 The deterministic variable $\varphi$ is a solution of
\Eq{evol},
and  $c(\bk)$ is a stochastic Gaussian field with power spectrum
\be
 \langle|c(\bk)|^2\rangle = (k/k_1)^{-2|\mu|-1} = (k/k_1)^{\alpha - 4} ~,
\ee
where we have related the tilt $\alpha$ introduced before to the
parameter $|\mu|$ used in \cite{1}.
 In order not 
to over-produce axions, we have to require $|\mu| \le 3/2$ i.e. $\alpha \ge 0$.
 The limiting value $\al=0$ corresponds precisely to a HZ
spectrum of CMB anisotropies on large scales~\cite{1}.

The energy momentum tensor of the axionic seeds is given by
\be
T_\mu^\nu=\dd_\mu\si\dd^\nu\si-{1\over 2}\de_\mu^\nu
\left(\dd_\al \si\right)^2~.
\label{Tsigma}
\ee
Like $\si$ also the energy momentum tensor is a stochastic variable
which is however not Gaussian. (The non-Gaussianity
of the model has to be computed and compared with observations.
But this is not the topic of the present work.)

For a universe with a given cosmic fluid, the linear perturbation
equations in Fourier space are of the form
\be
 \DD X = \SS~, \label{diff}
\ee
where $X$ is a long vector containing all the fluid perturbation
variables which depends on the wave number $\bk$ and conformal time
$\eta$. $\SS$ is a source vector which vanishes in the absence of
seeds. $\SS$ consists of linear
combinations of the seed energy momentum tensor and $\DD$ is a linear
ordinary differential operator.
More concretely, we consider a universe consisting of cold dark matter,
baryons, photons and three types of massless neutrino with a total
density parameter $\Om=1$, with or without 
 a cosmological constant ($\Omega_{\Lambda}=0.7$ or $0.0$). We choose the
baryonic density parameter $\Om_B=0.05$ and the value of the Hubble  
parameter  $H_0 =100h$km/sMpc with $h=0.5$.
More details on the linear system of
differential equations~(\ref{diff}) can be found in Ref.~\cite{DKM}
and references therein.

 Since $\SS$ is a stochastic variable,
so will be the solution $X(\eta_0)$ of \Eq{diff}.
We want to determine power spectra or, more generally, quadratic
expectation values of the form (with sums over repeated indices understood)
\be
\langle X_iX_j^*\rangle =\int_{\eta_{in}}^{\eta_0}
	\GG_{il}(\eta)\GG_{jm}^*(\eta')
	\langle \SS_l(\eta)\SS_m^*(\eta')\rangle d\eta d\eta' ~,
\label{power}
\ee
where $\GG$ is a Green's function for $\DD$.

We therefore have to compute the unequal time correlators,
$\langle \SS_l(\eta)\SS_m^*(\eta')\rangle$, of the seed energy momentum
tensor. This problem can, in general, be solved by an eigenvector
expansion method~\cite{Turok}.
 If the source evolution is linear, the problem becomes
particularly simple. In this 'coherent' case, we have
\[
 \SS_j(\eta) =f_{ji}(\eta,\eta_{in})\SS_i(\eta_{in})
\]
with a deterministic transfer function $f_{ij}$. 
By a simple change of variables we can diagonalize the hermitian,
positive initial equal time correlation  matrix, so that
$ \langle \SS_l(\eta_{in})\SS_m^*(\eta_{in})\rangle =\la_l\de_{lm} $.
Inserting this in \Eq{power} 
we obtain exactly the same result as by
replacing the stochastic variable $\SS_j$  by the deterministic source
 term $ S_j^{(det)}$ given by
\[
 S_j^{(det)}(\eta)S_i^{(det)*}(\eta')
	 =  \exp(\th_{ji})
 \sqrt{\langle |\SS_j(\eta)|^2\rangle\langle |\SS_i(\eta)|^2\rangle} ~,
\]
where the phase $\th_{ji}$ has to be
determined case by case. 

For our problem, the evolution of the pseudo-scalar field $\si$ is
linear, but the source, the energy momentum tensor of $\si$, is
quadratic in the field. The same situation is met for the large-$N$
approximation of global $O(N)$ models. There one finds that
the full incoherent result  is not very different from the perfectly  coherent 
approximation~\cite{DKM}. We hence are  confident that we obtain
relatively accurate results (to about 15\%) in the perfectly coherent 
approximation which we apply in our numerical calculation. A more
thorough discussion of the accuracy of the  coherent approximation
will be given in a forthcoming paper~\cite{VMDV}.
Within the coherent approximation, we just need to determine the equal
time correlators of the axion energy momentum tensor,
$
 \langle T_{\mu\nu}(\bk,\eta) T^*_{\rho\la}(\bk',\eta) \rangle
$,
which are fourth order in  $\si$. 

We then split the perturbations into scalar, vector, and tensor
parts which completely decouple within linear perturbation theory.

We determine the CMB anisotropies by numerically solving 
\Eq{evol}, and inserting the resulting source functions in a
Boltzmann solver.  

As discussed in \cite{1}, the amplitude of the CMB anisotropies  
depends on the small scale
cutoff, $k_1$, of the axion spectrum and the ratio between the string
scale $M_s$ and the Planck mass $M_P$ in the way
\be
 \ell(\ell+1)C_{\ell} \simeq (M_s/M_P)^4(\ell/k_1\eta_{dec})^{2\al} ~.
 \label{norm}
\ee
The simplest assumption,
$k_1/a_1\sim M_s \simeq 10^{-2}M_{P} \simeq 10^{17}{\rm GeV}$ only
leads to the correct normalization if $\al \ltapprox 0.1$. Otherwise the
 tilt factor $(k_1 \eta_{dec})^{- 2 \alpha}\sim 10^{-60 \alpha}$
entirely suppresses fluctuations on large scales. The huge
factor $k_1 \eta_{dec}$ comes from extrapolating the spectrum over 30 orders of
magnitude. If the tilt is larger than $\al\sim 0.1$, as suggested 
by the data (see below), we need either a
slightly scale dependent tilt or some cutoff in the 
small scale fluctuations at later times.  These possibilities are both
 physically plausible. The first one is realized if
the compactified dimensions evolve more rapidly
at the beginning of the dilaton-driven inflationary phase than towards its end.
In other words the parameter $r$ and $\al$ in Eq.(\ref{spectralindex}),
 instead of being constant,
will be a (slowly) decreasing function of time.
One could thus have a rather blue spectrum
at large scales, as necessary in order to have pronounced peaks, and a much
flatter spectrum at small scales which helps avoiding normalization
problems. We explore
these questions in more detail in the forthcoming paper~\cite{VMDV}.

\begin{figure}[ht]
\centerline{\epsfig{file=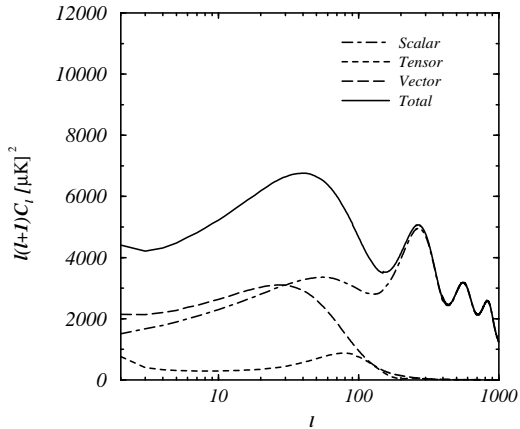, width=3.1in}}
\caption{The CMB anisotropy power spectrum for fluctuations induced 
from axionic seeds with a tilt $\al=0.1$. We show the 
scalar (dot-dashed), vector (dashed) and 
tensor (dotted) contributions separately as well as their sum 
(solid).}
\end{figure}

In Fig.~1 we show the scalar, vector and
tensor contributions to the resulting CMB anisotropies for an axion
spectrum with tilt $\al=0.1$. The 'hump' at $\ell\sim 40$ is due to
the isocurvature nature of the perturbations. They are also the main
reason why the acoustic peaks are very low. The result is remarkably
similar to the large-$N$ case studied in Ref.~\cite{DKM}. The main
difference here is that, like for usual inflationary models,  we 
dispose of a spectral index which is basically free. By
choosing slightly bluer spectra, we enhance the power on
smaller scales. 

\begin{figure}[ht]
\centerline{\epsfxsize=3.1in  \epsfbox{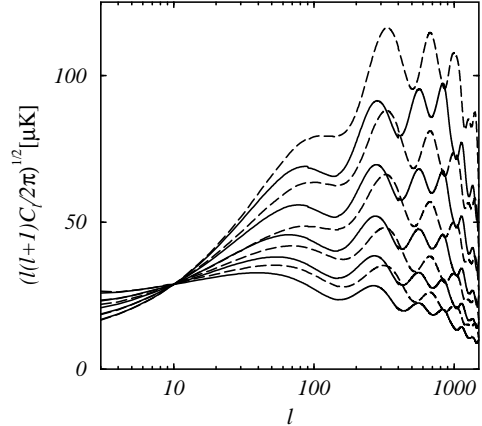}}
\caption{The CMB anisotropy power spectrum for fluctuations induced 
by axionic seeds. We show the sum of the scalar, vector  and 
tensor contributions for 5 different tilts, with 
$\Omega_{\Lambda}=0$ (solid) and $\Omega_{\Lambda}=0.7$
(long dashed). The tilt is raising from bottom to top,
 $\al=0.1,~0.2,~0.3,~0.4,~0.5$.}
\end{figure}

\begin{table}[ht]
\begin{center}
\begin{tabular}{|c|c|c|c|c|c|}
\hline
$\al =$ & 0.1 & 0.2 & 0.3 & 0.4 & 0.5 \\
\hline
 $\chi^2$ for $\La=0$ & 302  & 214 & 119  & 66 & 82\\
$\chi^2$ for $\La=0.7$ & 249  & 152 & 111 & 70 & 119 \\
\hline
\end{tabular}
\end{center}
\caption{The value of $\chi^2$ (with 15\% theoretical errors) from all
the CMB anisotropy experiments 
compiled in Refs.~\protect\cite{Teg} are presented for all the models. 
We compare with $N=60$ data points. 
Clearly, $\al \sim 0.4$ with $\La=0$ or $0.7$ is a
reasonable fit to the data.}
\end{table} 

In Fig.~2 we compare the results from different tilts
with and without cosmological constant. The CMB power spectra obtained 
can have considerable acoustic peaks at $\ell \sim 250$ to $300$.
Increasing the tilt $\al$ raises the acoustic peaks and moves them
slightly to smaller scales. 
As found in Ref.~\cite{1}, the power spectrum of the scalar component 
 is always blue.  The tensor and
 vector components counterbalance the increase of the tilt,
 maintaining a nearly scale invariant spectrum on large scales.
The models can be
 discriminated from the common inflationary spectra by their 
isocurvature hump and by the position of the first peak.
We have compared our results with
the latest experiments ~\cite{Teg}. All the models agree quite
well with the large scale experiments, while on degree and sub-degree
scales, models with $0.3 \ltapprox \al \ltapprox 0.5$ are favored by the data
as can be seen from the $\chi^2$ analysis presented in Table~I.
For comparison, the $\chi^2$ of a standard $\La$-CDM model, with 
theoretical errors given by cosmic variance, is $~120$. However, we
have to be aware that the $\chi^2$-test with present observations 
is a very rough indication of the goodness of a model, since the 
$C_\ell$s do not obey a Gaussian distribution\cite{Jaffe}.  This is 
especially serious for experiments with low sky coverage!

In Fig.~3, the theoretical dark matter power spectra are compared
with the data as compiled by Peacock and Dodds ~\cite{Pea}.
Models without a cosmological constant disagree in
shape and amplitude with the data. The root mean square mass fluctuation
within a ball of radius $8h^{-1}$Mpc for these models is 
$\sigma_8 = 0.36,0.56,0.88,1.36,2.05$
for the tilts from  $\al=0.1$ to $\al=0.5$ respectively.
Models with a cosmological constant are in reasonable agreement
with the shape of the spectrum (see Fig.~3). The values
of $\sigma_8$ for these models are $0.21, 0.38, 0.53, 0.82,1.25$ 
respectively. We estimate a (normalization) error of up to 
$\sim 30 \%$ in these numbers, due to
the perfectly coherent approximation. 
 Analysis of the abundance of galaxy clusters
suggest $\si_8 \sim 0.5(1-\Om_\La)^{-0.5}$~\cite{Eke}.
Since we can choose a blue, tilted spectrum in our model, we have more
power on small scales and are able to fit large scale structure data
much better than defect models for which the spectral index is fixed
by causality.

\begin{figure}[ht]
\centerline{\epsfxsize=3.1in  \epsfbox{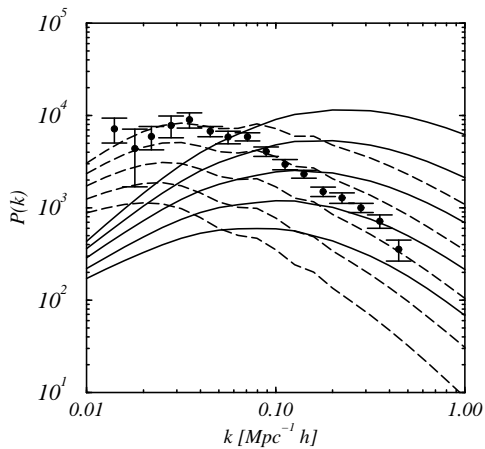}}
\caption{The linear dark matter power spectrum for fluctuations induced
by axionic seeds is compared with data for same values of the tilt as 
in Fig.~2. The spectra for the $\Om_\La \neq 0$ models (dashed lines) 
are shown with a bias factor of $b=1.2$. The value of the tilt raises
from bottom to top as in Fig.~2.
}
\end{figure}

In this {\it letter} we have presented preliminary results
for the CMB anisotropies and linear matter power spectra in a
pre-big bang scenario with axionic seeds. Due to the isocurvature
nature of the perturbations, a positive
tilt $0.3 \ltapprox \al \ltapprox 0.5$ is required to fit the 
measured CMB anisotropy. Including a
cosmological constant of $\Omega_{\Lambda} \sim 0.7$,
as suggested by the recent supernovae results~\cite{Perl}, the
matter power spectrum is also in good agreement with measurements.

If improved data confirms the need of a significant tilt, $\al>0.1$,
the most simple scenario ($k_1/a_1=M_s$ and $\al=$ const.) will be ruled out.
This shows that CMB anisotropies may contain information about the
evolution of extra dimensions! But clearly, also in this case 
the model remains highly predictive. It is  easily distinguished from
the more standard adiabatic models by its 'isocurvature hump' at $\ell
< 100$ and the position of the first acoustic peak at $\ell \sim
300$. These values depend only slightly on the tilt (see Fig.~2).
Furthermore the ratios between the scalar, vector and tensor contributions
are entirely fixed by the model. 

We are grateful to Ram Brustein for helpful comments and to Nicola
 Vittorio for his Boltzmann code. This work is
 supported by the Swiss NSF.


\begin{thebibliography}{99}
\bibitem{MAP}See the web sites:~ {\tt http://map.gsfc.nasa.gov}, and
  {\tt http://astro.estec.esa.nl/SA-general/ \\
\qquad Projects/Planck} 
\bibitem{PBB} G.~Veneziano, Phys. Lett. B  {\bf 265}, 287 (1991);
	M. Gasperini and G. Veneziano,
	 Astropart. Phys. {\bf 1}, 317 (1993).
An updated collection of papers is available at 
  {\tt http://www.to.infn.it/\~{}gasperin}.
\bibitem{Copeland}E.J. Copeland, R. Easther, and D. Wands,
	Phys. Rev. D {\bf 56}, 874 (1997).
\bibitem{Buon}
A. Buonanno, {\it et al.}, JHEP01 (1998)  004.
\bibitem{Hadad}
R. Brustein and M. Hadad, \prd {\bf 57} (1998) 725.
\bibitem{d90}  R. Durrer, { Phys. Rev. } D {\bf 42}, 2533 (1990);
	 R. Durrer, { Fund. of Cosmic Physics} {\bf 15}, 209
	(1994).
\bibitem{1} R. Durrer, M. Gasperini, M. Sakellariadou and G.
	Veneziano, \prd {\bf 59}, 043511 (1999);
	Phys. Lett. {\bf B436}, 66 (1998).
\bibitem{3} M. Gasperini and G. Veneziano, \prd {\bf 59}, 043503 (1999).
\bibitem{isocurvature} P.J.E. Peebles,  { Astrophys. J.} {\bf 510}, 523
	(1999); {\em ibid.} 531. 
\bibitem{COBE}G.F. Smoot {\em et al.}, { Astrophys. J.} {\bf 396}, L1 (1992);
	 C.L. Bennett {\em et al.}, { Astrophys. J.} {\bf 430}, 
	423 (1994).
\bibitem{Teg} M. Tegmark,~{\tt http://www.sns.ias.edu/\~{}max};
  	A.D. Miller et al., preprint {\tt astro-ph/9906421} (1999).
\bibitem{Turok} N. Turok,  Phys. Rev. D {\bf 54}, 3686 (1996).
\bibitem{DKM} R. Durrer, M. Kunz, and A. Melchiorri, Phys. Rev. D {\bf 59} 
	123005 (1999).
\bibitem{VMDV} F. Vernizzi {\em et al.}, in  preparation (1999).
\bibitem{Jaffe}J.R. Bond, A.H. Jaffe \& L. Knox,  Astrophys. J., to appear
	 (1999, {\tt astro-ph/9808264})
\bibitem{Pea} J. Peacock and S. Dodds, Mon. Not. R. Astron. Soc. {\bf 267},
	1020 (1994).
\bibitem{Eke} V. Eke, S. Cole, and C. Frenk  Mon. Not. R. Astron. Soc. {\bf
	282}, 263 (1996).
\bibitem{Perl} S. Perlmutter {\em et al.}, Astrophys. J., in print (1999). 
\end{thebibliography}
\end{document}